\documentclass{article}

\usepackage[T1]{fontenc}
\usepackage[utf8]{inputenc}
\usepackage[english]{babel}
\usepackage{geometry}
\usepackage{lipsum}
\usepackage{tikz}

\usepackage{booktabs}

\usepackage[english]{varioref}
\usepackage{amsmath}
\usepackage{amsthm}
\usepackage{amsfonts}
\usepackage{graphicx}
\usepackage{sidecap}
\usepackage[font=small,hang]{caption}
\usepackage{subfig}

\usepackage{url}
\usepackage{hyperref}
\usepackage{CormorantGaramond}

\usepackage{multicol}
\usepackage{lipsum}% for dummy text

\begin{document}
\title{How do simple evolutionary strategies and investment optimizations affect ecological patterns? The case of generalized Taylor's Law}
\author{Stefano Garlaschi \footnote{Department of Physics and Astronomy "Galileo Galilei'', University of Padova} and Samuele Stivanello \footnote{Department of Mathematics "Tullio Levi-Civita'', University of Padova} }
\date{}
\maketitle
\begin{abstract}
\noindent

\textbf{Taylor’s Law (TL) relates the variance to the mean of a random variable via power law. In ecology it applies to populations and it is a common empirical pattern shared among different ecosystems. Measurements give power law exponent to be between 1 and 2, and more often to cluster around 2, whereas  theoretical models predict TL exponent can assume any real value. In this paper, adopting the framework of multiplicative growth models in a Markovian environment, we investigate the possibility of evolutionary strategies  to be responsible for TL exponent to be in a finite range. We implement three different strategies the individuals can follow and for each strategy set two different optimization investment objectives. In all the studied cases we find  TL exponent can assume any real value due to the existence of regions of the model parameters in which the exponent can diverge. Furthermore, under natural hypothesis on the dynamics of the environment, the shapes of these regions do not depend on different strategies adopted and nor on the optimization objective. Thus the introduction of strategies dose not affect the range of TL exponent in the model. In our theoretical framework rare events are shaping the value of the TL exponent, suggesting, as hinted by previous works, that empirical values may be a statistical artifact following from under sampling.}

\end{abstract}

\begin{multicols}{2}

\section{Introduction}
%citare articolo originale TL
Taylor's Law (TL) \cite{Taylor} is a deeply investigated pattern in ecological dynamics which states that for a population size  $N$ the variance scales like a power law of its mean:
\begin{equation}
\text{Var[N]}\simeq a \mathbb{E}[N]^b
\end{equation}
with $a>0$ and $b \in \mathbb{R}$.

This statistical regularity can be extended to the cumulants of $N$ and we refer to this as generalized TL, which relates via power law the $k$th cumulant to the $j$th one:
\begin{equation}
\mathbb{E}[N^k]\simeq a_{jk}\mathbb{E}[N^j]^{b_{jk}}.
\end{equation}

%citare vari lavori presenti in PNAS, in particolare magari quello del data set della black rock forest
TL and generalized TL have been largely tested across different ecosystems (\cite{TL-test}-\cite{TL-test-2})and their validity becomes more and more corroborated while collecting new data. In particular it emerges that $b_{jk}$ is close to the ratio of the order of the two cumulants in relation, $b_{jk} \simeq k/j$, hence in the common version of Taylor's Law $b$ assumes only bounded values near $2$. TL has been proved to be stable also with different inter-individuals dynamics: researchers showed how TL and its exponent is not remarkably affected by the introduction of a mechanism of competition for the food sources between two bacteria cultures (\cite{competition}). Another feature is worth to mention is how both of these regularities emerge also in different field, such as physical sciences or human dynamics studies.

%citare vari lavori teorici
In order to try to uncover the existence of an underlying generating mechanisms of these recurrent patterns and to predict the range of values of the exponents, a lot of theoretical works started to be undertaken (\cite{theo-work-1}-\cite{last-theo-work}). An example of those is \cite{PNAS} in which a multiplicative growth model in a Markovian environment (\cite{A}, \cite{B}) was built. %With this the population size at time $t$ is written as

Using Large Deviations Theory (LDT) techniques (\cite{denhollander}-\cite{LDT-MC})to study the model, they obtained different results  a disagreement with what is naturally observed. 

As a consequence of this disagreement, a natural question arises: is $b_{jk} \simeq k/j$ the result of some ecological processes that their model didn't consider or is it a pure statistical artifact? 
%ha senso spiegare il loro lavoro in un articolo? alla fine uno se gli interessa è supposto andare a vederlselo per conto suo...

This latter possibility may be due to an undersampling in the measurement procedure which is not able to detect the so called rare events (i.e. events with low probability of realization) that instead the model takes into account and in the same paper the authors suggested this scenario.

In this paper we set under investigation the influence of different evolutionary strategies adopted by the individuals on the generalized TL.

\section{The strategies of the model}

We build up a multiplicative growth model for a population living in a Markovian environment
\begin{equation}
N(t)  = N_0  \prod_{n=1}^t \tilde{A}_n
\label{eq:pop-size}
\end{equation}
%in which the environment is modelled by the same Markov chain $A_{t,t\geq 0}$. 
In details, we are supposing the environment to be able to acquire two different states, let us call these $0$ and $1$, and its evolution is described by a Markov chain with the symmetric transition matrix
\begin{equation}
P=
\begin{pmatrix}
1-\lambda && \lambda \\
\lambda && 1-\lambda
\end{pmatrix}
\end{equation}
with $0<\lambda <1$. So $\lambda$ represents the probability that the environments changes state in two consecutive steps.

%So we modeled the environment with the chain ${A_{t,t\geq 0}}$ with space state $\Gamma=\{0,1\}$. 
Now the individuals can adopt two phenotypes, again labeled $0$ and $1$, each one adapted to a different environment state. Individuals with phenotype $0$ (resp. 1) fits better in environment $0$ (resp. 1), meaning that they grow
by a factor $r$ (it is natural to consider $r>1$); whereas if they are in the unsuitable environment 1 (resp. 0) they decrease by a factor $s$ (naturally $0<s<1$).

In our model we introduce a strategy, meaning that at every time step $n$ each individual decides to adopt the phenotype $S_{n+1}$. The strategy succeeds if $S_{n+1} (A_n) = A_{n+1}$, i.e. if the individual chooses its phenotype according to the next-step environment basing its decision on the environment state it is seeing.

\subsection{The adaptive strategy}

In the first situation we considered is the one in which all individuals play the same adaptive strategy, i.e. $S_{n+1}(A_n)=A_n$. In this way we describe a population which tries to adapt itself to the current environment.

%Adaptive in the sense that the phenotype adopted at the step $n$, let us call this $S_n$, is the one good for the environment at time $n-1$, i.e. $S_n=A_{n-1}$. The situation describes a population which tries to become adapt to the environment until this changes.

With this strategy the multiplicative factor $\tilde{A}_n$ is
\[
\tilde{A}_n=\left[r\cdot \delta_{S_{n}, A_{n}}+s\cdot \left(1-\delta_{S_{n}, A_{n}}\right)\right]=
\]
\begin{equation}
=\begin{cases}
r \qquad \text{if } S_{n} = A_{n-1} = A_{n} \\
s \qquad \text{if } S_{n} =A_{n-1} \neq A_{n}
\end{cases}
\end{equation}

Let us introduce the empirical pair measure $\overline{L_t^2}$ defined as 
\begin{equation}
\overline{L_t^2}=\frac{1}{t}\sum_{n=1}^t \delta_{A_{n-1},A_n}
\end{equation} 
counting the fraction of times the chain does not change state in a realization of the Markov chain up to time $t$.  It is easy to demonstrate that the family $P_t(\mu)=\mathbb{P}(\overline{L_t^2} \in [\mu,\mu+d\mu])$ satisfies a Large Deviations Principle (theorem $\text{IV}.3$ and theorem $\text{III}.5$, \cite{denhollander}) with rate $t$ and rate function
\begin{equation}
I(\mu)=\mu \log \left(\frac{\mu}{1-\lambda}\right)+(1-\mu)\log \left(\frac{1-\mu}{\lambda}\right).
\end{equation} 
where $\mu$ $\left(\mu \in [0,1]\right)$ is the proportion of times the chain does not change state in a realization of the Markov chain up to time $t$.

So it follows that the population size \eqref{eq:pop-size} can be written as
\begin{equation}
N(t)=N_0 e^{tG(\mu)},
\end{equation}
where 
\begin{equation}
G(\mu)=\mu \log r + (1-\mu)\log s.
\end{equation}

%$\mu \in [0,1]$ is defined as the fraction of times the chain does not change state, i.e.
%\begin{equation}
%\mu=\frac{1}{t}\sum_{n=1}^t \delta_{A_{n-1},A_n}
%\end{equation}
%and 

%The quantity $\mu$, since depends on the specific chain realization, can be framed mathematically in the context of the Probability Theory as an empirical random measure. It is easy to demonstrate that $\mu$ satisfies a LDP (theorem $\text{IV}.3$ and theorem $\text{III}.5$, \cite{denhollander}) with rate $t$ and rate function
%\begin{equation}
%I(\mu)=\mu \log \left(\frac{\mu}{1-\lambda}\right)+(1-\mu)\log \left(\frac{1-\mu}{\lambda}\right).
%\end{equation}
%We can see that this rate function has the right properties (it is non-negative and convex $\forall \mu \in [0,1]$ and it has a absolute minimum).

In the following we will set $N_0=1$ without any loss of generality and we will focus on the first and the second cumulant. So for simplicity of notation we will refer to $b_{12}$ as $b$. Clearly

%To compute this quantity we consider the ratio between $t^{-1} \log \mathbb{E}\left[N(t)\right]$ and $t^{-1} \log \mathbb{E}\left[N^2(t)\right]$ as discussed in \cite{Cohen}. For the moments of the random variable $N(t)$ we can apply Varadhan's Lemma which gives us
%\begin{equation}
%\lim_{t \to +\infty} t^{-1} \log \mathbb{E}\left[N(t)^k\right]= \sup_{\mu \in [0,1]}\left[kG(\mu)-I(\mu)\right].
%\end{equation}
%and so
%\begin{equation}
%b(\lambda)=\frac{\sup_{\mu \in [0,1]}\left[2G(\mu)-I(\mu)\right]}{\sup_{\mu \in [0,1]}\left[G(\mu)-I(\mu)\right]}
%\label{eq:exp-GTL-general-form}
%\end{equation}

\begin{equation}
b(\lambda)=\frac{ t^{-1}\log\mathbb{E}\left[N(t)^2\right]}{ t^{-1}\log\mathbb{E}\left[N(t)\right]}
\end{equation}

Here we adopt LDT, applying Varadhan's Lemma (theorem $\text{III}.13$, \cite{denhollander}), which states
\begin{equation}
\lim_{t \to +\infty} t^{-1} \log \mathbb{E}\left[N(t)^k\right]= \sup_{\mu \in [0,1]}\left[kG(\mu)-I(\mu)\right].
\end{equation}

In this way we have
\begin{equation}
b(\lambda)=\frac{\lim_{t \to +\infty} t^{-1}\log\mathbb{E}\left[N(t)^2\right]}{\lim_{t \to +\infty} t^{-1}\log\mathbb{E}\left[N(t)\right]}=\frac{\sup_{\mu \in [0,1]}\left[2G(\mu)-I(\mu)\right]}{\sup_{\mu \in [0,1]}\left[G(\mu)-I(\mu)\right]}
\label{eq:exp-GTL-general-form}
\end{equation}

Inserting the expression of the two functions and computing the two suprema we find $b$ as a function of $\lambda$ parametrized by $r$ and $s$:
\begin{equation}
b(\lambda)=\frac{\log\left[(1-\lambda)r^2+\lambda s^2\right]}{\log\left[(1-\lambda
)r+\lambda s\right]}
\label{eq:exp_single_strat}
\end{equation} 

\begin{figure*}
\centering
\includegraphics[scale=0.5]{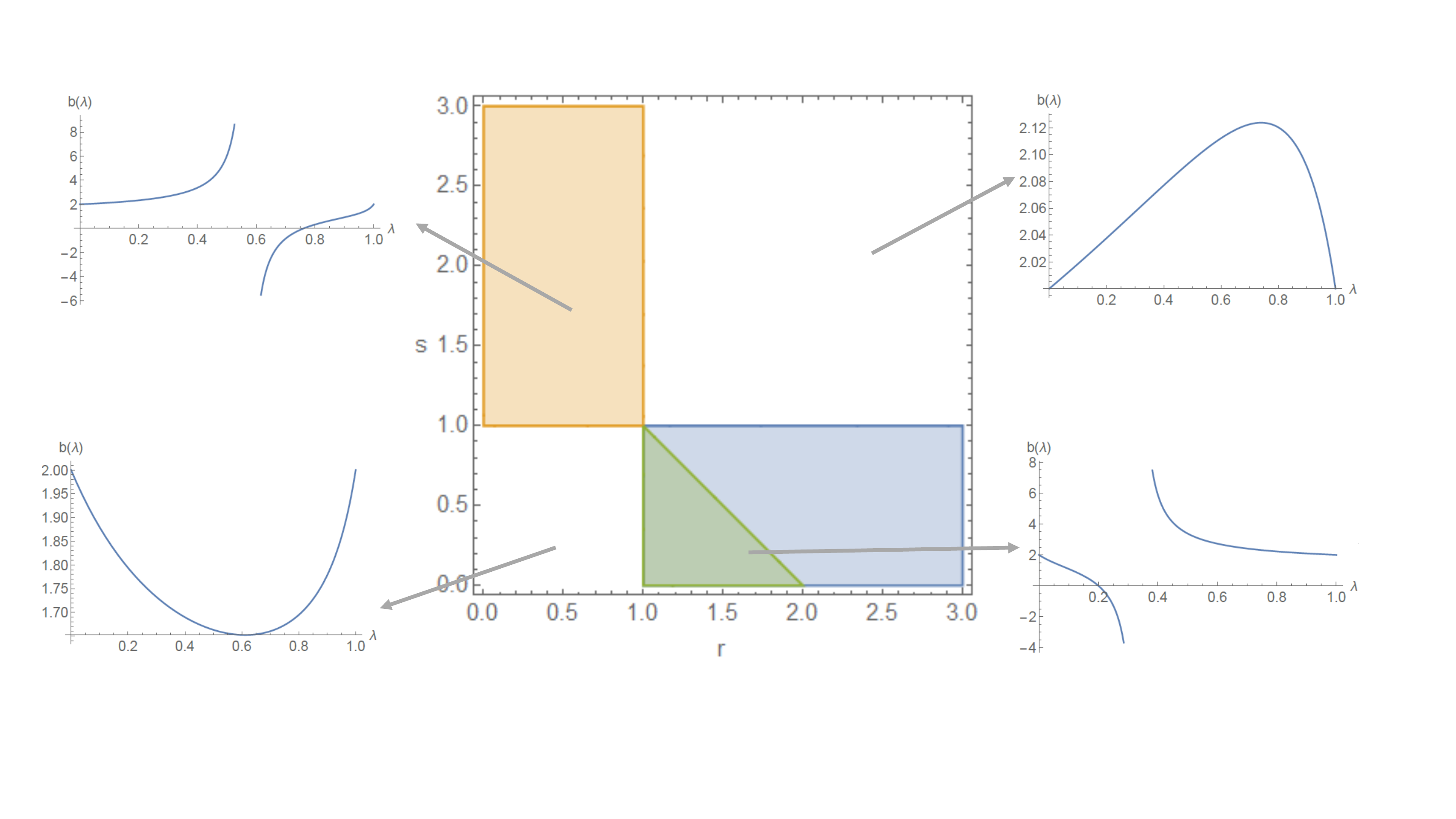}
\caption{Regions of divergence for the exponent of the generalized TL with the adaptive strategy. The small panels show $b(\lambda)$ for $r$ and $s$ in the different zones. Only in the coloured regions $\lambda_c \in [0,1]$ and so the exponent diverges: in blue is shown $R_1$ which contains the sub-region $\overline{R_1}$ in green, while in orange $R_2$ is shown.}
\label{fig:div_single_strat}
\end{figure*}

We can see that Eq. \eqref{eq:exp_single_strat} may diverge for a critical value $\lambda_c$. Searching when the denominator becomes zero we find
\begin{equation}
\lambda_c=\frac{r-1}{r-s}
\end{equation}
The divergence shows up only when $\lambda_c \in [0,1]$ and so in the regions
\begin{equation}
\begin{cases}
R_1=\{(r,s):0<s<1, r>1\} \\
R_2=\{(r,s):0<r<1, s>1\}
\end{cases}
\end{equation}
The situation is shown in Figure \ref{fig:div_single_strat}. As we can see in the whole "natural" region of the parameters ($r>1$ and $0<s<1$) the exponent displays a divergence. Although, we have to consider that this adaptive strategy can be meaningful only if the environment does not change frequently, i.e. the probability of changing is smaller than the probability of remaining. So we have to consider in our model $0<\lambda<\frac{1}{2}$. 

Now the subregion of divergence $\overline{R_1} \subset R_1$ in which $0<\lambda_c<\frac{1}{2}$ is considerably smaller since it is defined by
\begin{equation}
\overline{R_1}=\left\lbrace (r,s):r>1,0<s<1,r+s<2\right\rbrace
\end{equation}
This region is also shown in Figure \ref{fig:div_single_strat}.

\subsection{Mixed strategy}

Secondly we introduce a new strategy line in the model. Now a fraction $p$ of the population adopts the adaptive strategy described above, but the other $1-p$ fraction use the completely opposite strategy, that consists in betting on the change of the
environment, i.e. $S_{n+1}(A_n)=1-A_n$. 
%These individuals bet on the changing of the environment and so at the next step they use the phenotype which was not good for the previous environment. Thus while $S_{p,n}=A_{n-1}$, for this new strategy $S_{1-p,n}=1-A_{n-1}$.

For this reason the multiplicative factor now is
\[
\tilde{A}_n=\left[\overline{r}\cdot \delta_{A_{n-1}, A_{n}}+\overline{s}\cdot \left(1-\delta_{A_{n-1}, A_{n}}\right)\right]=
\]
\begin{equation}
=\begin{cases}
\overline{r} \qquad \text{if } A_{n-1} = A_{n} \\
\overline{s} \qquad \text{if } A_{n-1} \neq A_{n}
\end{cases}
\end{equation}
with
\begin{equation}
\begin{cases}
\overline{r}=rp+(1-p)s\\
\overline{s}=r(1-p)+ps
\end{cases}
\label{eq:rinomina-rs}
\end{equation}
From this dynamics we obtain for $N(t)$ 
\begin{equation}
N(t)=N_0 e^{tG(\mu)}
\end{equation}
where $\mu$ is the same quantity as above.
% (so we will use the same rate function), because also there is no distinction between the transitions, but the only thing that matters is if the environment changed. Instead the function $G(\mu)$ is changed
We will use the same rate function $I(\mu)$ since we are again interested in counting the fraction of times the environment changes; but note that the different strategy has effect on the rate of growth, i.e.
\begin{equation}
G(\mu)=\mu \log \overline{r}+(1-\mu)\log \overline{s}.
\end{equation}

So now the exponent, starting from Eq. \eqref{eq:exp-GTL-general-form} again, becomes
\begin{equation}
b(\lambda)=\frac{\log\left[(1-\lambda)\overline{r}^2+\lambda \overline{s}^2\right]}{\log\left[(1-\lambda
)\overline{r}+\lambda\overline{s}\right]}.
\label{eq:exp_two_strat}
\end{equation}

The denominator can still becomes zero for
\begin{equation}
\lambda_c=\frac{\overline{r}-1}{\overline{r}-\overline{s}}.
\end{equation}
and in the following we will investigate if it is possible that $\lambda_c \in [0,1]$.

Until now we have considered $p$ as a parameter. We wonder ourselves: is there a natural choice of $p$? Ecologically a population tries to maximize its growth and so we look for the optimal value $p^*$ maximizing the rate of growth of the population on the long term. We can do this under two different considerations:
\begin{itemize}
\item accounting only the most probable realizations of the chain and so replacing $\mu $ with the value determined by the Law of Large Numbers (LLN);
\item considering also the contribution of the rare events and so using Large Deviation Theory (LDT) techniques.
\end{itemize}

\subsubsection{Optimizing with LLN}
\begin{figure*}
\centering
\subfloat[][$r=4$ and $s=0.7$.]
   {\includegraphics[width=.4\textwidth]{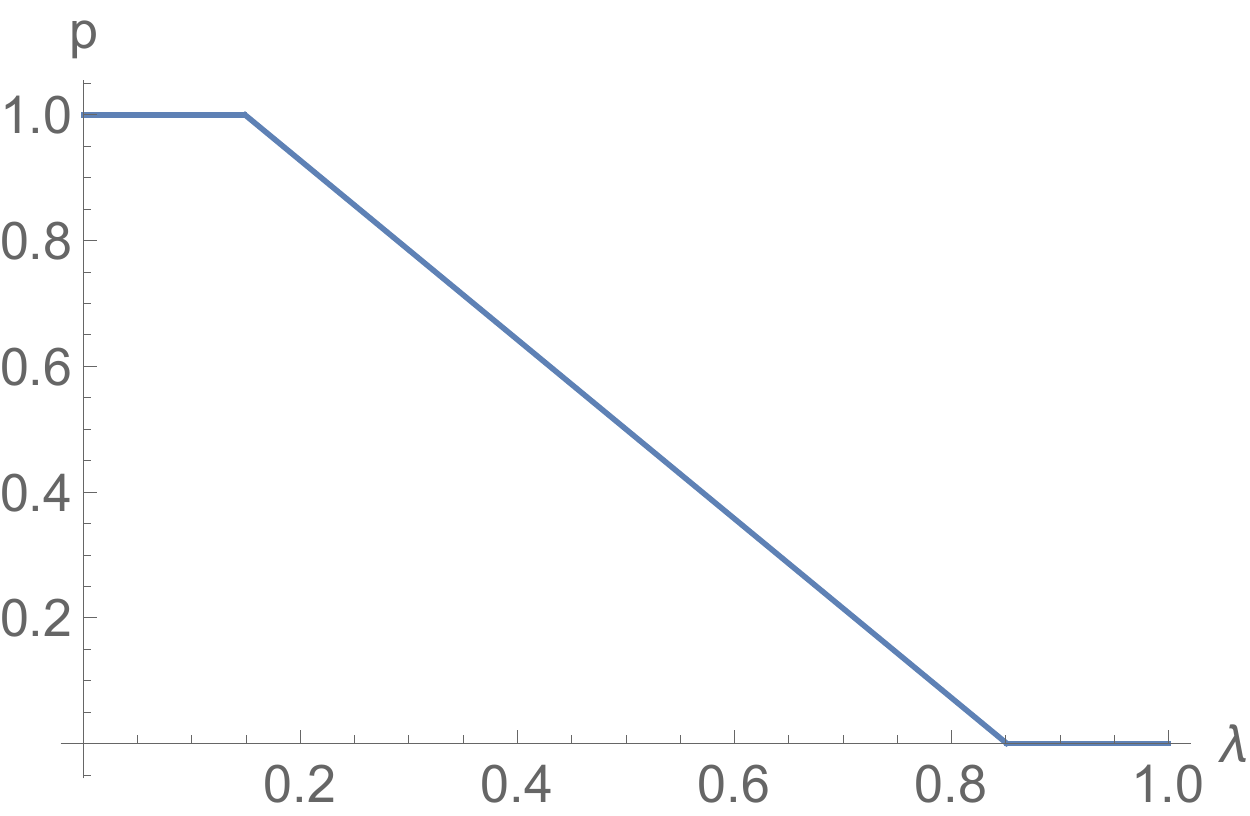}} \quad
\subfloat[][$r=0.4$ and $s=1.6$.]
   {\includegraphics[width=.4\textwidth]{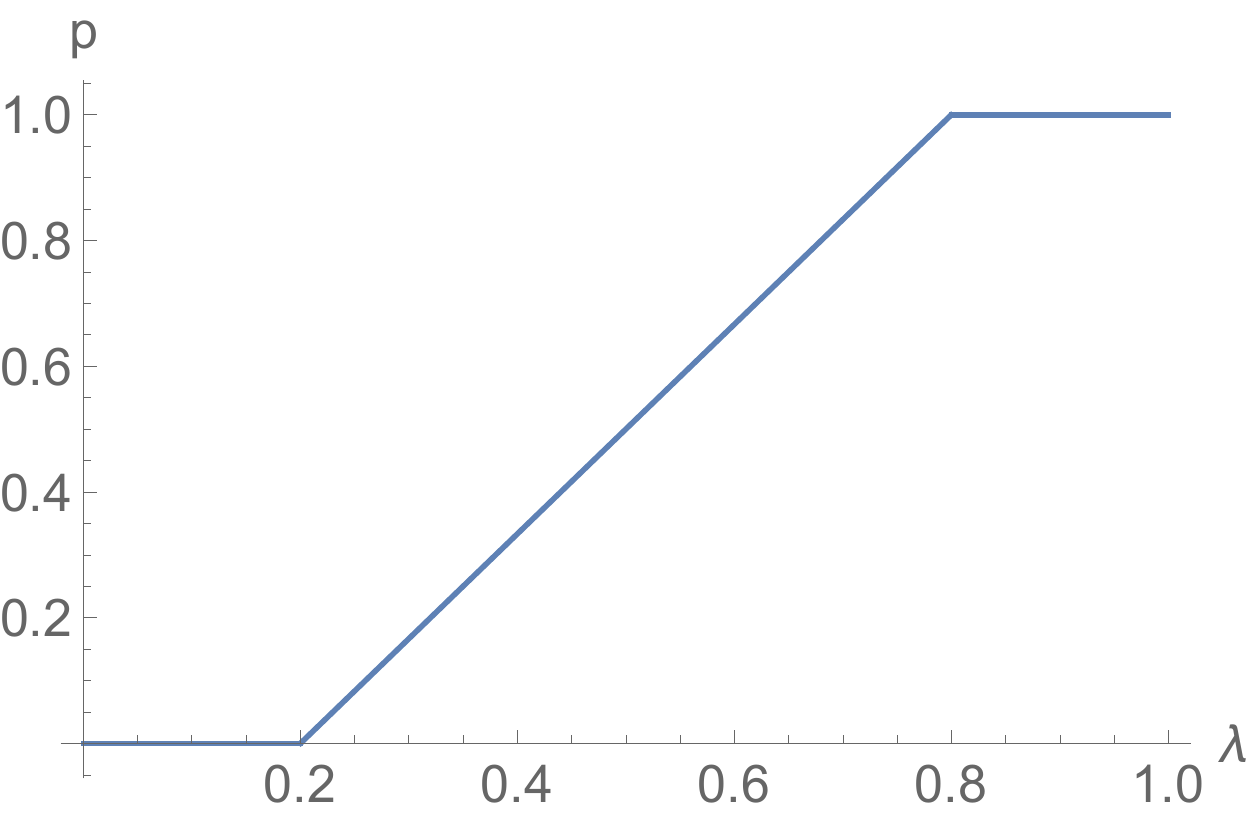}} \\
\caption{Plots of $p^{*}(r,s,\lambda)$ for different values of $r$ and $s$.}
\label{fig:p_LLN_two_Strat}
\end{figure*}

In the first case we apply LLN. This state that $\mathbb{P}(\mu = 1-\lambda) \to_{t \to +\infty} 1$. So we have to maximize with respect to $p \in [0,1]$ the quantity
\begin{equation}
G(\mu=1-\lambda)=(1-\lambda) \log \overline{r}+ \lambda \log \overline{s}
\end{equation}
obtaining 
\begin{equation}
p^{*}(r,s,\lambda)=
\begin{cases}
1 \qquad \qquad \hspace{0.4 cm} \text{if } 0 < \lambda \leq \frac{s}{r+s} \\
\frac{r(1-\lambda)-\lambda s}{r-s} \hspace{0.5 cm} \text{if } \frac{s}{r+s} \leq \lambda \leq \frac{r}{r+s} \\
0 \qquad \qquad \hspace{0.4 cm}\text{if } \frac{r}{r+s}\leq \lambda <1
\end{cases}
\end{equation}
if $r>s$ or
\begin{equation}
p^{*}(r,s,\lambda)=
\begin{cases}
0 \qquad \qquad \hspace{0.4 cm} \text{if } 0 < \lambda \leq \frac{r}{r+s} \\
\frac{r(1-\lambda)-\lambda s}{r-s} \hspace{0.5 cm} \text{if } \frac{r}{r+s} \leq \lambda \leq \frac{s}{r+s} \\
1 \qquad \qquad \hspace{0.4 cm}\text{if } \frac{s}{r+s}\leq \lambda <1
\end{cases}
\end{equation}
if $r<s$. Two plots of these are shown in Figure \ref{fig:p_LLN_two_Strat}. As we can see for $r>s$, increasing $\lambda$, there is a gradual transition from a situation in which all the population uses the adaptive strategy to a situation in which nobody adopts it passing through a coexistence of the two (and viceversa for $r<s$).

With this expression for $p^{*}$ Eq. \eqref{eq:exp_two_strat} for $b(\lambda)$ becomes a function of $\lambda$ parametrized by $r$ and $s$. 

Also in this case $b$ can displays a discontinuity. The new critical value of $\lambda_c$ now is a function of $\lambda$ due to its dependence on $p^{*}$. So there is a divergence when 
\begin{equation}
\lambda=\lambda_c(\lambda) \hspace{0.2 cm} \text{with } \lambda \in [0,1].
\end{equation}
Solving this equation we find four possible regions. These are four since there is a first division depending on $r>s$ or $r<s$ which give two different $p^*$; then there is a second ramification due to the possibility of having $\lambda_c$ in the intervals of values of $\lambda$ in which $p=0$ or $p=1$ or in the interval of coexistence of the two strategies, i.e. $0<p<1$. These regions are   
\begin{itemize}
\item In the first region $r>1$ and $0<s<1$ and the divergences are in correspondence of the intervals of $\lambda$ in which $p^{*}=0$ or $p^{*}=1$. So
\[
R_1=\lbrace(r,s):\sqrt{-4 s^2+4 s+1}-2 r>-1,
\]
\[
0<s<1, r>1\rbrace
\]
and
\[
\lambda_c=\frac{r-1}{r-s} \vee \lambda_c=\frac{1-s}{r-s}
\]
\item In the second $r>1$ and $0<s<1$ and the divergences are in correspondence of the interval of $\lambda$ in which $0<p^{*}<1$. So
\[
R_2=\{(r,s):\sqrt{-4 s^2+4 s+1}-2 r<-1,
\]
\[
r+s<2,0<s<1, r>1\}
\]
and
\[
\lambda_c=\frac{1}{2}\pm\frac{1}{2} \sqrt{\frac{2-r-s}{r+s}}.
\]
\item In the third region $0<r<1$ and $s>1$ and the divergences are in correspondence of the interval of $\lambda$ in which $p^{*}=0$ or $p^{*}=1$. So
\[
R_3=\lbrace(r,s):\sqrt{-4 r^2+4 r+1}-2 s>-1,
\]
\[
0<r<1, s>1\rbrace
\]
and 
\[
\lambda_c=\frac{r-1}{r-s} \vee \lambda_c=\frac{1-s}{r-s}.
\]
\item In the last region $0<r<1$ and $s>1$ and the divergences are in correspondence of the interval of $\lambda$ in which $0<p^{*}<1$. So
\[
R_4=\{(r,s):\sqrt{-4 r^2+4 r+1}-2 s<-1,
\]
\[
r+s<2, 0<r<1, s>1\}
\]
and
\[
\lambda_c=\frac{1}{2}\pm\frac{1}{2} \sqrt{\frac{2-r-s}{r+s}}.
\]
\end{itemize}

\begin{figure*}
\centering
\includegraphics[scale=0.5]{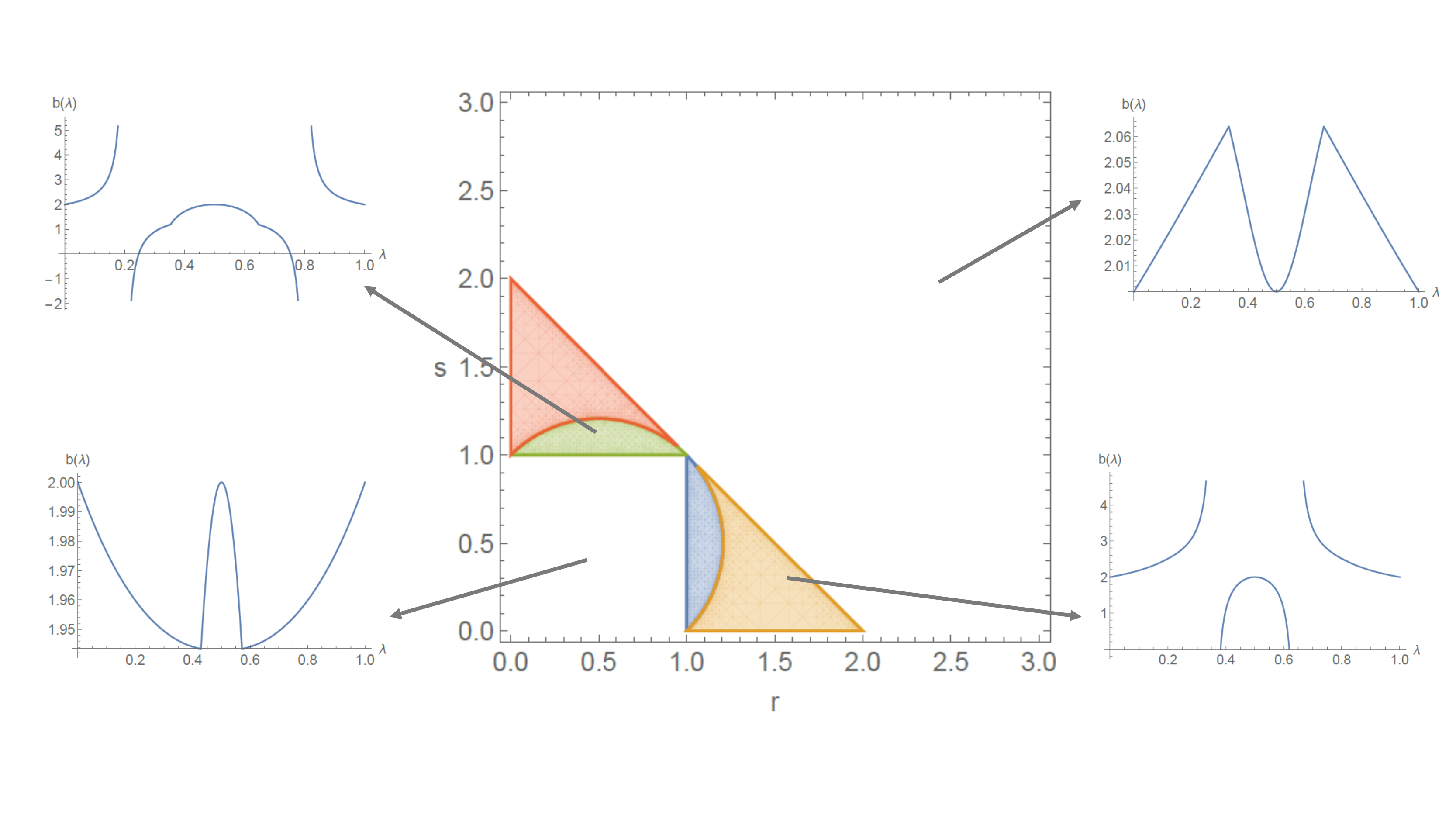}
\caption{Regions of divergence for the exponent of the generalized TL with the mixed strategy optimized using LLN. The blue zone is $R1$, the orange one is $R2$, the green zone represents $R3$ and the red zone is $R4$. The small panels show $b(\lambda)$ for $r$ and $s$ in the different zones. }
\label{fig:div_LLN}
\end{figure*}

In all of these regions the exponents displays two divergences symmetric with respect to $\frac{1}{2}$ and  hence there is always a $\lambda_c < \frac{1}{2}$. The Figure \ref{fig:div_LLN} shows the four region of divergence. 

The region of divergence shrank in a sensible way with respect to the adaptive strategy alone. But this positive effect disappears when we take into account only the "natural" region: in this case we obtain the same divergence-region as in the previous model.
%As it is possible to see now the situation is much improved but a residual region of divergence is still present. In the "natural" region we have the same region of divergence looking for $\lambda_c < \frac{1}{2}$ as in the previous model.

\subsubsection{Optimizing accounting rare events}

In order to take account also rare events we have to consider all the contribute from all the possible values of $\mu$. Thus we have to maximize $\mathbb{E} \left[ N(t) \right]$. Taking the logarithm (since it is a monotone function) and looking on the long term (i.e. $t \to +\infty$) we can exploit Varadhan's Lemma and so we have to maximize the quantity
\begin{equation}
\log\left[(1-\lambda)\overline{r}+\lambda\overline{s}\right]
\end{equation}
with respect to $p$.

In this way we find for $r>s$
\begin{equation}
p^{*}(\lambda,r,s)=
\begin{cases}
1 \quad \text{if } \lambda<\frac{1}{2} \\
0 \quad \text{if } \lambda\geq\frac{1}{2}
\end{cases}
\end{equation}
and for $r<s$
\begin{equation}
p^{*}(\lambda,r,s)=
\begin{cases}
0 \quad \text{if } \lambda<\frac{1}{2} \\
1 \quad \text{if } \lambda\geq\frac{1}{2}
\end{cases}
\end{equation}

Investigating the region of divergence we find
\begin{itemize}
\item $\widetilde{R_1}=\{(r,s):1 < r < 2 - s \wedge 0< s < 1 \}$ with 
\item $\widetilde{R_2}=\{(r,s):1 < r < 2 - s \wedge 0< s < 1 \}$.
\end{itemize}
and in both cases we have
\begin{equation}
\lambda_c=\frac{r-1}{r-s} \vee \lambda_c=\frac{1-s}{r-s}.
\end{equation}
Also in this case we have two symmetric $\lambda_c$.

\begin{figure*}
\centering
\includegraphics[scale=0.5]{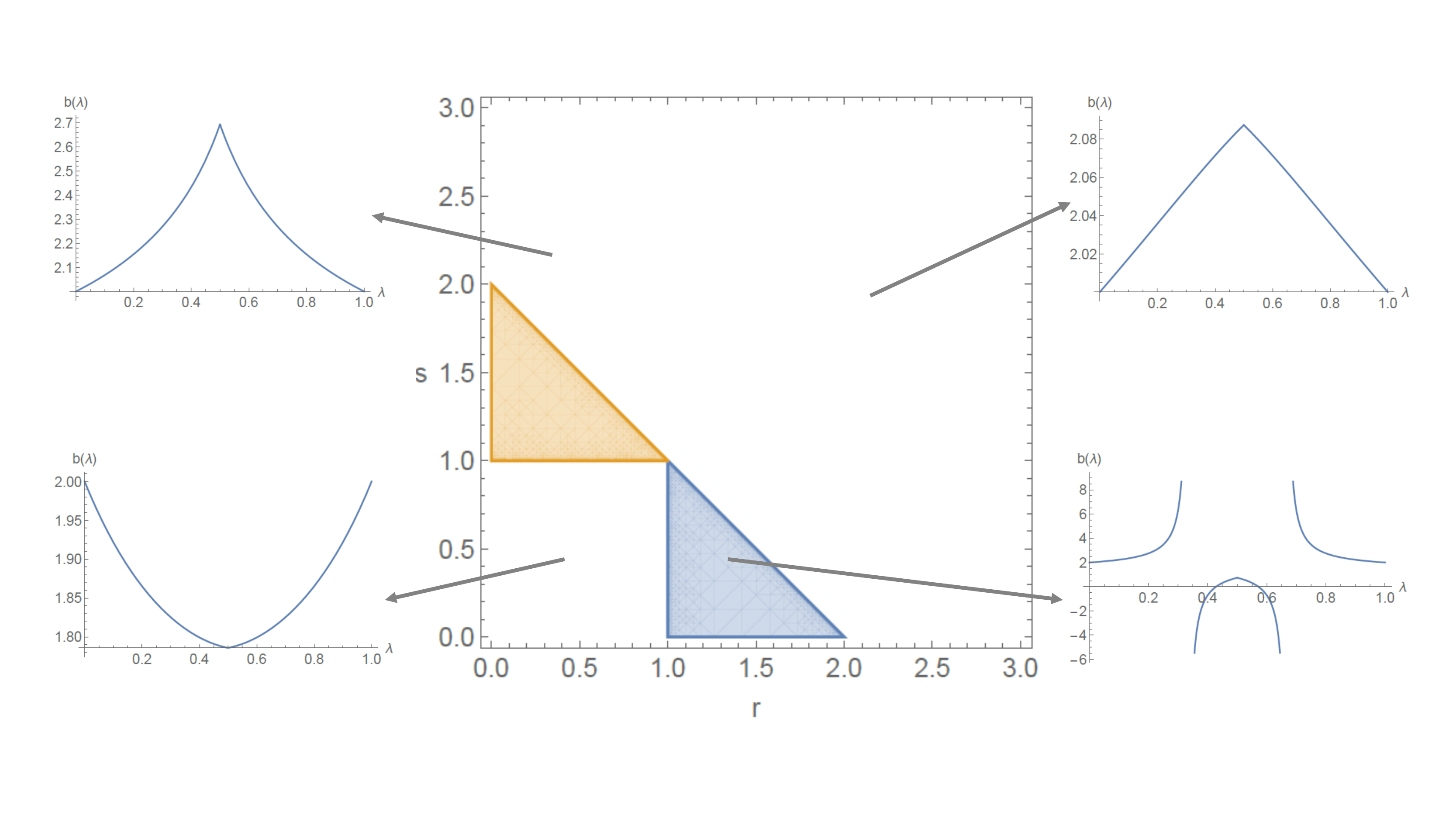}
\caption{Regions of divergence for the exponent of the generalized TL with the mixed strategy optimized using LDP to take trace even of rare events. The blue zone represents $\widetilde{R_1}$. The orange one instead is $\widetilde{R_2}$. The small panels display some plots of $b(\lambda)$ for $r$ and $s$ in the different regions.}
\label{fig:div_LDP}
\end{figure*}

We can see these regions in Figure \ref{fig:div_LDP}.

\subsection{Adaptive strategy with the presence of a randomly moving fraction}

The last situation we analyzed consists in a population of which a fraction $p$ adopt the adaptive strategy and the other part $1-p$ moves randomly, i.e. this fraction has a fifty-fifty probability of choosing one of the two phenotypes for the step $n$ without taking care of what the environment was at time $n-1$. An easy calculation based on the transition matrix $P$ and on the steady distribution of the chain (since we will look for large $t$) shows how this random motion is equivalent to remain still in one phenotype, i.e. using always the same (let us say $0$ without any loss of generality). So $S_{p,n}=A_{n-1}$ and $S_{1-p,n}=0$.

In the previous situations we had only to count the fraction of times the environment did or did not change without taking care which states were involved. What mattered was only the changing or the persistence of the environment. Now instead we can not do the same any more since the growth now it is not determined only by the fraction of times the environment change.

So now we have to consider each possible transition of the chain separately. With this in mind we can find

\begin{equation}
\tilde{A}_n=
\begin{cases}
r \quad \text{if } (A_{n-1},A_n)=(0,0) \\
s \quad \text{if } (A_{n-1},A_n)=(0,1) \\
\overline{s} \quad \text{if } (A_{n-1},A_n)=(1,0) \\
\overline{r} \quad \text{if } (A_{n-1},A_n)=(1,1)
\end{cases}
\end{equation}

As we can see, each transition gives a different multiplicative factor. Now the population at time $t$ can be written as
\begin{equation}
N(t)=N_0 e^{tG(\nu)}
\end{equation}
where $\nu$ is a four-components object
\begin{equation}
\nu=(\nu_{ij})_{i,j=0,1}
\end{equation}
that counts the fraction of times each transition happened, i.e.
\begin{equation}
\nu_{ij}=\frac{1}{t}\sum_{n=1}^t \delta_{(A_{n-1},A_n),(i,j)}
\end{equation}
with the obvious constraint
\begin{equation}
\sum_{i,j=0}^1 \nu_{ij}=1.
\end{equation}
So $\nu \in \widetilde{\mathfrak{M}}_1\left(\Gamma \times \Gamma\right)$.

Now
\begin{equation}
G(\nu)=\nu_{00} \log r+ \nu_{01} \log s+ \nu_{10} \log \overline{s} + \nu_{11} \log \overline{r}.
\end{equation}

The domain of $\nu$ is
\begin{equation}
\Delta^4=\left\lbrace(\nu_{ij})_{i,j=0,1}:0\leq \nu_{ij} \leq 1, \sum_{i,j=0}^1 \nu_{ij}=1\right\rbrace.
\end{equation} 

Let us introduce the empirical pair measure $L_t^2$ defined as
\begin{equation}
L_t^2=\frac{1}{t}\sum_{n=1}^t \delta_{(A_{n-1},A_n),}
\end{equation}
counting the fraction of times each possible transition occurs in a realization of the Markov chain up to time $t$. As demonstrated in \cite{denhollander} the family $P_t(\nu)=\mathbb{P}\left(L_t^t \in [\nu,\nu+d\nu]\right)$ satisfies a LDP with rate $t$ and rate function
%This new quantity $\nu$ is a measure on $\Gamma \times \Gamma$ and so $\nu \in \widetilde{\mathfrak{M}}_1\left(\Gamma \times \Gamma\right)$. As demonstrated in \cite{denhollander} $\nu$ satisfies a LDP with rate function
\begin{equation}
I^2_P(\nu)=\sum_{i,j}\nu_{ij}\log\left(\frac{\nu_{ij}}{\overline{\nu}_i P_{ij}}\right)
\end{equation}
with $\overline{\nu}_i=\sum_t \nu_{ij}$ and $i,j \in \Gamma$.

Having the rate function and the function $G(\nu)$ we can compute the exponent 
\begin{equation}
b(\lambda)=\frac{\sup_{\nu \in \Delta^4}\left[2G(\nu)-I^2_P(\nu)\right]}{\sup_{\nu \in \Delta^4}\left[G(\nu)-I^2_P(\nu)\right]}
\end{equation}
Unfortunately these two suprema cannot be carried out analytically and so it is necessary to implement a numerical procedure to compute them and the exponent. Again, we still have to fix $p$ with an optimal value that maximizes the growth rate.

\subsubsection{Optimizing with LLN}

Applying LLN it holds
\begin{equation}
\nu=\left(\frac{1-\lambda}{2},\frac{\lambda}{2},\frac{\lambda}{2},\frac{1-\lambda}{2}\right).
\end{equation}
With this result we look for the maximum of the function $G\left(\frac{1-\lambda}{2},\frac{\lambda}{2},\frac{\lambda}{2},\frac{1-\lambda}{2}\right)$ with respect to $p$ finding the same $p^{*}$ of the mixed strategy with LLN.

\begin{figure*}
\centering
\includegraphics[scale=0.5]{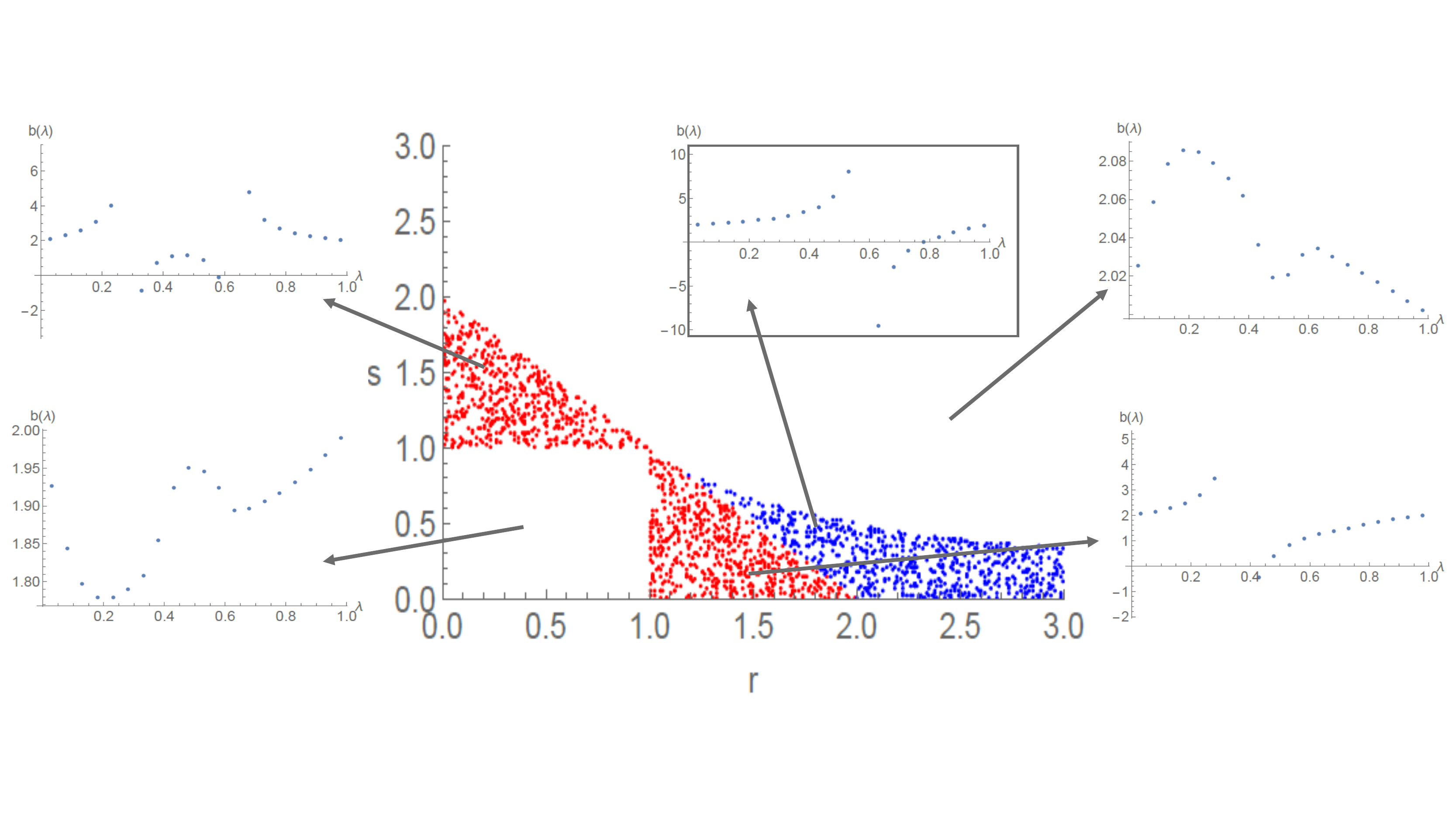}
\caption{Regions in which the exponent $b(\lambda)$ diverges for the situation with fraction using adaptive strategy or randomness in the choice of the phenotype optimized by LLN. To reconstruct the regions points , due to the absence of analytical results, an uniform sampling of the region $[0,3] \times [0,3]$ was performed, then the exponent is compute for different $\lambda$. In this way we looked if the exponent diverge. The blue points make $b$ diverges with $\lambda_c>\frac{1}{2}$. The red one instead provide $\lambda_c<\frac{1}{2}$. In this way we derived numerical evidences on what the region could be. The small panels shows some plots obtained via numerical calculation of the exponent for a couple $r$ and $s$ in the different regions of the parameters obtained in such a way.}
\label{fig:div_LLN_wr}
\end{figure*}

To numerically compute the two suprema we fix $r$, $s$ and $\lambda$. In this way we get also $p=p^{*}(r,s,\lambda)$. Now we get the two suprema and taking the ratio we obtain the exponent. Repeating this for different $\lambda$ we can piece together the plot of $b(\lambda)$ and investigate the existence of critical values $\lambda_c$. The results of this way of calculation are graphically displayed in Figure \ref{fig:div_LLN_wr}. As we can see, again we find a region of divergence. Splitting the region to underline where $\lambda_c < \frac{1}{2}$, we find the same sub region as in the previous situations. 

\subsubsection{Optimizing accounting rare events}

With this optimization we maximize $\log\mathbb{E} \left[ N(t) \right]$ as done above. To do that previously we exploited Varadhan's Lemma and we can do it again here. So we have to maximize
\begin{equation}
\sup_{\nu \in \Delta^4}\left[G(\nu)-I^2_P(\nu)\right]
\end{equation}
We have to do this numerically. We fix $r$, $s$ and $\lambda$ and we compute this supremum for different value of $p$. The optimal value $p^{*}$ for these $r$, $s$ and $\lambda$ is the one giving the greatest numerical value for the supremum. Changing $\lambda$ we obtain $p^{*}$ as a function of $\lambda$ with the parameters $r$ and $s$ fixed. 

Numerically evidences provide the same $p^{*}$ obtained in the case of mixed strategy maximized with LDP. 

\begin{figure*}
\centering
\includegraphics[scale=0.5]{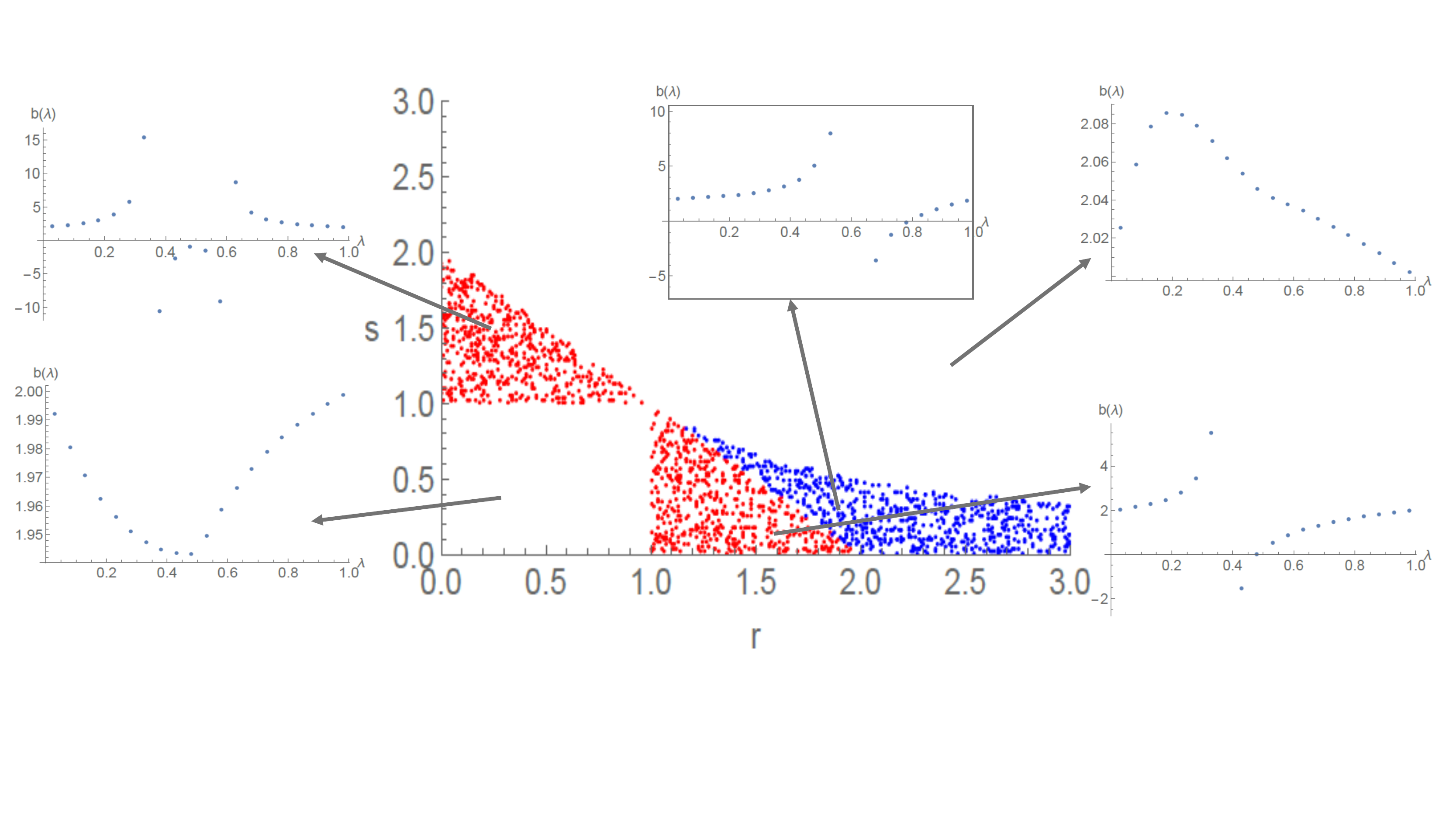}
\caption{Regions in which the exponent $b(\lambda)$ diverges for the situation with fraction using adaptive strategy or randomness in the choice of the phenotype optimized using LDT. To reconstruct the regions points , due to the absence of analytical results, an uniform sampling of the region $[0,3] \times [0,3]$ was performed, then the exponent is compute for different $\lambda$. In this way we looked if the exponent diverge. The blue points make $b$ diverges with $\lambda_c>\frac{1}{2}$. The red one instead provide $\lambda_c<\frac{1}{2}$. In this way we derived numerical evidences on what the region could be. The small panels shows some plots obtained via numerical calculation of the exponent for a couple $r$ and $s$ in the different regions of the parameters obtained in such a way.}
\label{fig:div_LDP_wr}
\end{figure*}

With $p^*$ found in this way we can compute the other supremum and taking the ratio we obtain the value of the exponent for that $\lambda$. The results are shown in Figure \ref{fig:div_LDP_wr}. With this optimization we find the same regions.

\section{Conclusions} 

We implemented a model in which the population can perform some evolutionary strategies, i.e., based on the current environment, betting on the right phenotype to adopt for the future environment. We investigate the possibility of evolutionary strategies to be the mechanisms that generate TL and generalized TL and make the exponents displaying only bound values in nature.

In the model we used tools of Large Deviation Theory and we studied the possibility that rare events do not play a important role for the right estimation of the exponent. Instead we find, as in \cite{PNAS}, the exponent can diverge, i.e. it can assume any real value. Looking for the region in which $\lambda_c < \frac{1}{2}$ in all the considered situations we found the same region independently on the strategy adopted or the optimization performed.

So we can conclude that even introducing different strategies in the model rare events may play a fundamental role for the right computation of the exponent without any relevance on dynamics details. We think that the limited range of values of the exponent that are observed in nature could still be due to an undersampling of the real world dynamics. Such a conclusion is also suggested in \cite{PNAS} 

We hope that this primer work can inspire future research in this direction. Many routes are traced. Does an optimal strategy exists? This optimal strategy should be characterized by the smallest divergence-region. Can we optimize p in a different way?
The divergence on $b(\lambda)$ appears when the denominator goes to zero. This can be seen as the population approximates extinction . Here fluctuations of the population size play a fundamental role: if these are big enough, then extinction can occur. So a possible optimization is to search for a $p$ minimizing fluctuations but allowing a growth of the population on the long term.
%We need to fix a treshold $S$ and look for $\tilde{p}$ that maximize the growth rate under the constraint that fluctuation are smaller than $S$.

\end{multicols}

\end{document}